\documentclass[sigconf]{acmart}
\usepackage{multirow} 

\AtBeginDocument{%
  }

\setcopyright{acmlicensed}
\copyrightyear{2025}
\acmYear{2025}
\acmDOI{XXXXXXX.XXXXXXX}
\acmConference[Conference acronym 'XX]{Make sure to enter the correct
  conference title from your rights confirmation email}{June 03--05,
  2018}{Woodstock, NY}
\acmISBN{978-1-4503-XXXX-X/2018/06}

\begin{document}

\title{UniShare: A Unified Framework for Joint Video and Receiver Recommendation in Social Sharing}

\author{Caimeng Wang}
\authornote{Both authors contributed equally to this research.}
\affiliation{%
  \institution{Kuaishou Technology}
  \city{Beijing}
  \country{China}
}
\email{wangcaimeng@kuaishou.com}

\author{Li Chong}
\authornotemark[1]
\affiliation{%
  \institution{Kuaishou Technology}
  \city{Beijing}
  \country{China}
}
\email{chongli@kuaishou.com}

\author{Dongxu Liu}
\affiliation{%
  \institution{Kuaishou Technology}
  \city{Beijing}
  \country{China}
}
\email{liudongxu06@kuaishou.com}

\author{Xu Min}
\authornote{Corresponding author}
\affiliation{%
  \institution{Kuaishou Technology}
  \city{Beijing}
  \country{China}
}
\email{minxu@kuaishou.com}

\author{Jianhui Bu}
\affiliation{%
  \institution{Kuaishou Technology}
  \city{Beijing}
  \country{China}
}
\email{bujianhui@kuaishou.com}

\renewcommand{\shortauthors}{Wang et al.}

\begin{abstract}
Sharing behavior on short-video platforms constitutes a complex ternary interaction among the user (sharer), the video (content), and the receiver. Traditional industrial solutions often decouple this into two independent tasks: video recommendation (predicting share probability) and receiver recommendation (predicting whom to share with), leading to suboptimal performance due to isolated modeling and inadequate information utilization. To address this, we propose \textbf{UniShare}, a novel unified framework for joint sharing prediction on both video and receiver recommendation. UniShare models the share probability through an enhanced representation learning module that incorporates pre-trained GNN and multi-modal embeddings, alongside explicit bilateral interest and relationship matching. A key innovation is our joint training paradigm, which leverages signals from both tasks to mutually enhance each other, mitigating data sparsity and improving bilateral satisfaction. We also introduce \textbf{K-Share}, a large-scale real-world dataset constructed from Kuaishou platform logs to support research in this domain. Extensive offline experiments demonstrate that UniShare significantly outperforms strong baselines on both tasks. Furthermore, online A/B testing on the Kuaishou platform confirms its effectiveness, achieving significant improvements in key metrics including the number of shares (+1.95\%) and receiver reply rate (+0.482\%).

\end{abstract}



\received{20 February 2007}
\received[revised]{12 March 2009}
\received[accepted]{5 June 2009}

\maketitle

\section{Introduction}
\begin{figure}[t]
  \centering
\includegraphics[width=\linewidth]{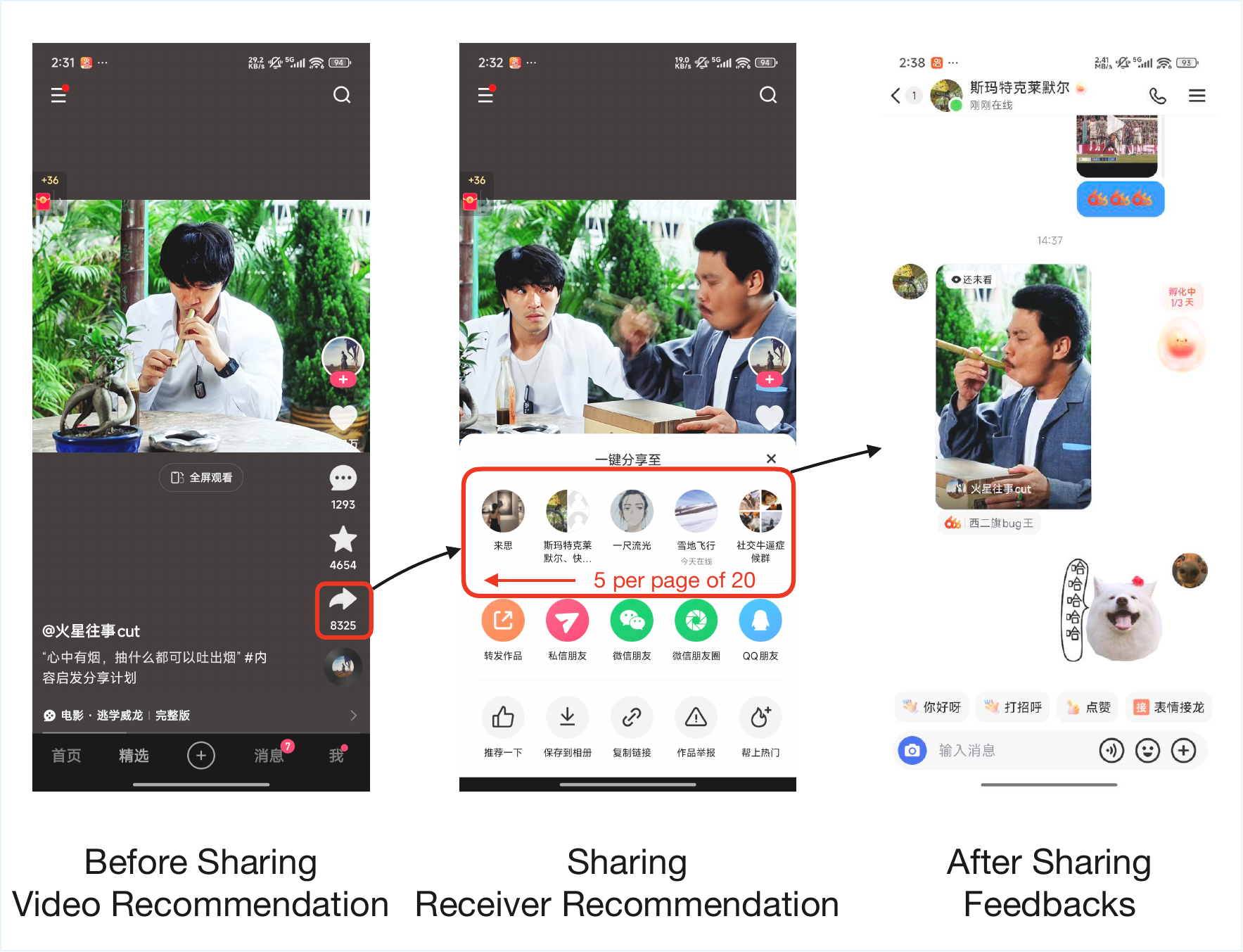}
  \caption{A screenshot illustrating the sharing flow on the Kuaishou platform. The process consists of three stages: (1) watching a video and generating sharing intent and clicking the share button to invoke the sharing panel, and (2) selecting a receiver to complete the sharing action, followed by (3) the receiver viewing the video and providing feedbacks.}
  \label{fig:intro}
\end{figure}
Social sharing has become a fundamental interaction mechanism on short-video platforms, forming a complex ternary relationship between the user (sharer), the video (content), and the receiver. This behavior is not merely a simple action but a core social activity that drives content dissemination and user engagement. As illustrated in Figure \ref{fig:intro}, the typical sharing flow involves: (1) a user watching a video, generating sharing intent, and clicking the share button; (2) selecting a specific receiver from a candidate list; and (3) the receiver consuming the shared content and potentially providing feedback.

Traditional industrial recommendation solutions often decouple this holistic process into two independent tasks: video recommendation (predicting the probability a user will share a video) and receiver recommendation (predicting whom the user will share with). Such decoupled approaches model these tasks in isolation, leading to several critical limitations: (1) Isolated Modeling: The video recommendation model typically lacks fine-grained receiver-side signals, while the receiver recommendation model often operates without leveraging deep semantic information from the specific video content being shared. (2) Inadequate Information Utilization: The rich, multi-modal signals and complex social dynamics inherent in the sharing triad are underutilized. (3) Data Sparsity: Sharing actions are inherently sparse compared to other in-app behaviors (e.g., clicks, views), making it challenging for separate models to learn effective representations, particularly for long-tail videos or infrequent sharing users. Consequently, these approaches fail to fully capture the bilateral satisfaction—simultaneously meeting the sharing intent of the initiator and the consumption preferences of the receiver—which is paramount for a successful sharing experience.

To address these challenges, we propose \textbf{UniShare}, a novel unified framework for joint video and receiver recommendation in social sharing. Unlike prior works, UniShare jointly models the probability of a successful share action through an enhanced representation learning module. This module incorporates pre-trained Graph Neural Network (GNN) embeddings and multi-modal features to alleviate sparsity, alongside explicit mechanisms for bilateral interest matching and relationship-content alignment. A key innovation of our framework is its joint training paradigm, which allows signals from both sub-tasks to mutually enhance each other, effectively mitigating data sparsity and improving overall recommendation performance for both videos and receivers.

Furthermore, to support research in this nascent area, we construct and release \textbf{K-Share}, a large-scale real-world dataset built from Kuaishou platform logs. K-Share comprehensively captures sharing behaviors, user social graphs, and multi-modal content features, providing a valuable testbed for evaluating ternary interaction recommendation models.

Extensive offline experiments on K-Share demonstrate that UniShare significantly outperforms strong separate baselines on both recommendation tasks. Moreover, online A/B testing conducted on the Kuaishou platform confirms the practical effectiveness of UniShare, showcasing significant improvements in key online metrics, including the number of shares (+1.95\%) and the receiver reply rate (+0.482\%).

The main contributions of this work are summarized as follows:
\begin{itemize}
    \item We identify the critical limitation of decoupled modeling for social sharing recommendation and formalize it as a joint prediction problem over the user-video-receiver triplet.
    \item We propose UniShare, a unified framework that integrates with explicit bilateral interest modeling and a novel joint training paradigm with hierarchical negative sampling, enabling mutual enhancement between the video and receiver recommendation tasks and effectively alleviating data sparsity.
    \item We introduce K-Share, a large-scale real-world dataset from Kuaishou platform to benchmark sharing prediction problems.  We plan to open-source the K-Share dataset upon completion of the release process.
    \item  We empirically validate the superiority of UniShare through rigorous offline experiments and demonstrate its significant practical value via successful online deployment and A/B testing.
\end{itemize}

\section{ Related Work}

\subsection{Joint Modeling in Recommendation}
Traditional deep recommendation systems (DRS) often rely on single-task, single-scenario, or single-modality data (e.g., ID-based features), which limits their ability to capture users’ complex and dynamic preferences. These methods suffer from narrow information extraction, high resource consumption, and underutilization of multi-source information. To address these issues, \textbf{Joint Modeling} has been introduced, integrating multi-dimensional information to improve recommendation accuracy, efficiency, and personalization \cite{Joint-Model-Survay}.

Many joint modeling methods have been proposed in recent years to handle multi-task and multi-scenario settings while mitigating negative transfer. For example, MMOE\cite{MMOE} uses a mixture-of-experts architecture with task-specific gating networks to share information across tasks while preserving task-specific features. STAR\cite{STAR} introduces a star-shaped topology with a shared center network and scenario-specific networks, effectively balancing common knowledge and scenario-specific characteristics. It laid the foundation for hard-sharing approaches in multi-scenario recommendation. SNR\cite{SNR}, PLE\cite{PLE}, and LHUC\cite{LHUC} also offer innovative structures for parameter sharing and customization.

Another line of work integrates auxiliary behaviors (e.g., clicks) to enhance target behavior prediction (e.g., purchase). ESMM\cite{ESMM} proposes a full-space multi-task model using Bayesian decomposition of CTR and CVR, inspiring successors such as ESM2\cite{ESM2}, DBMTL\cite{DBMTL}, GMSL\cite{GMSL} and EHCF\cite{EHCF} further explore task dependencies and hierarchical behavior modeling. There are also many multi-objective and multitasking optimization strategies emerging, such GradNorm\cite{GradNorm} , DWA\cite{DWA} , DTP\cite{DTP} , PCGrad\cite{PCGrad} and GradVac\cite{GradVec} . Additionally, MBGCN\cite{MBGCN} utilizes graph neural networks to implicitly learn behavior relationships, demonstrating the power of GNNs in multi-behavior recommendation.

\subsection{Ternary Interactions Modeling in Recommendation}
User-Item-Target (UIT) triplet modeling represents a core task in recommendation systems, social e-commerce, and social network analysis, aiming to model complex interactions among users, items, and targets. Distinguished from traditional dyadic modeling (e.g., User-Item, User-User), UIT simultaneously captures user preferences for items, item influence on target users, and synergistic three-way relationships, offering enhanced practicality at the cost of increased complexity.

HGSRec\cite{HGSRec} models ternary interactions among user-item-friend relationships via a triple heterogeneous graph neural network and transitive triplet representations, pioneering the combination of GNNs with transitive triplet representations for social e-commerce. ReSeq\cite{ReSeq} proposes a dual-perspective dynamic sequence modeling framework that employs Transformer to encode active/passive behavior sequences, combined with multi-scale matching prediction and self-distillation technology for efficient bidirectional preference modeling. DynShare\cite{DynShare} leverages bidirectional continuous-time dynamic graphs and a time-interval-aware projector to address asymmetric interactions and user inactivity in dynamic share recommendation. DPGNN\cite{DPGNN} constructs a dual-perspective interaction graph and hybrid preference propagation mechanism to explicitly model bidirectional selection preferences in job-person matching.

Social recommendation approaches further extend these concepts. SocialLGN\cite{SocialLGN} utilizes latent user interest information from social relationships as auxiliary information to alleviate data sparsity, extending LightGCN to social recommendation scenarios. Social-RippleNet\cite{Social-RippleNet} innovatively integrates both knowledge graph information and social network information to address sparsity and cold-start problems. DGNN\cite{DGNN} proposes a memory-enhanced disentangled graph neural network that explicitly models heterogeneous relationships (user-user, user-item, item-item) and their underlying latent factors, employing an innovative message propagation mechanism to automatically integrate disentangled semantic signals for enhanced social recommendation performance.

\section{Problem Formulation}
\begin{table}
  \caption{Notations.}
  \label{tab:notations}
  \begin{tabular}{cc}
    \toprule
    \textbf{Notation} & \textbf{Meaning} \\
    \midrule
    $\mathcal{U}$ & User universe ($U \in \mathcal{U}$) \\
    $\mathcal{I}$ & Video universe ($I \in \mathcal{I}$) \\
    $\mathcal{V}$ & Candidate receiver pool ($V \in \mathcal{V}$) \\
    $S$ & $S(U, I, V) = 
        \begin{cases} 
            1 & \text{share occurs} \\
            0 & \text{otherwise}
        \end{cases}$ \\
    $\mathcal{V}_u \subseteq \mathcal{V}$ & Valid receivers for specific user $u$ \\
    $u \in \mathcal{U}$ & A specific user. \\        
    $i \in \mathcal{I}$ & A specific video item.  \\
    $v \in \mathcal{V}$ & A specific receiver. \\  
    $\mathcal{H}$ & The historical behavior sequence \\
    $e$ & The pre-trained embedding \\
    $\mathcal{L}$ & The loss function \\
  \bottomrule
\end{tabular}
\end{table}

In social networks and content platforms, sharing behavior represents a fundamental form of social interaction. This research aims to address the prediction of sharing probability within ternary interactions involving users, videos, and receivers. Specifically, we formalize the problem as estimating the probability of a user initiating a successful sharing action given an input triplet $\langle U, I, V \rangle$, which consists of the following core components: a sharing initiator (user $U$), a candidate video (content $I$), and a candidate receiver (user $V$). 

The problem naturally decomposes into two interrelated sub-tasks that reflect real-world user behavior on social platforms: 
(1) \textbf{Video Share Probability Estimation}: For a fixed user $U$ and their candidate receiver set $\mathcal{V}_u$, this task requires estimating the probability that video $I$ will be shared by $U$ to any receiver within $\mathcal{V}_u$.
(2) \textbf{Receiver Recommendation}: For a fixed user $U$ and a specific video $I$, this task involves estimating a probability distribution over all candidate receivers $v \in \mathcal{V}_u $, representing the likelihood that $U$ will choose each receiver for sharing video $I$.

The core challenge lies in effectively modeling the ternary relationship through the conditional probability \begin{equation}
P(S=1 \mid U, I, V)
\end{equation}
where $S$ is a binary random variable indicating whether the sharing action occurs ($S$ = 1) or not. 

Mathematically, we define the two sub-tasks as follows:

\paragraph{Sub-task 1: User-Pair Video Share Probability Estimation}
For a fixed user $u \in \mathcal{U} $ and their candidate receiver set $\mathcal{V}_u$, we estimate the share probability for each candidate video $i \in \mathcal{I} $:
\begin{equation}
  f_1(i; u, \mathcal{V}_u) = \sum_{ \forall v \in \mathcal{V}_u}P(S=1 \mid U=u, I=i, V=v), \quad \forall i \in \mathcal{I}
\end{equation}

\paragraph{Sub-task 2: Per-Video Candidate Receiver Probability Estimation}  
For a fixed user $u \in \mathcal{U}$ and a specific video $i \in \mathcal{I} $, we estimate the probability distribution over all candidate receivers $v \in \mathcal{V}_u $:
\begin{equation}
    f_2(v; u, i) = P(S=1 \mid U=u, I=i, V=v), \quad \forall v \in \mathcal{V}_u
\end{equation}

Both sub-tasks derive from a unified probabilistic model that can be expressed as:
\begin{equation}
P(S=1 \mid U, I, V) = g(\mathbf{u}, \mathbf{i}, \mathbf{v}; \theta)
\end{equation}

\section{Methodology}
\begin{figure*}[h]
  \centering
  \includegraphics[width=\linewidth]{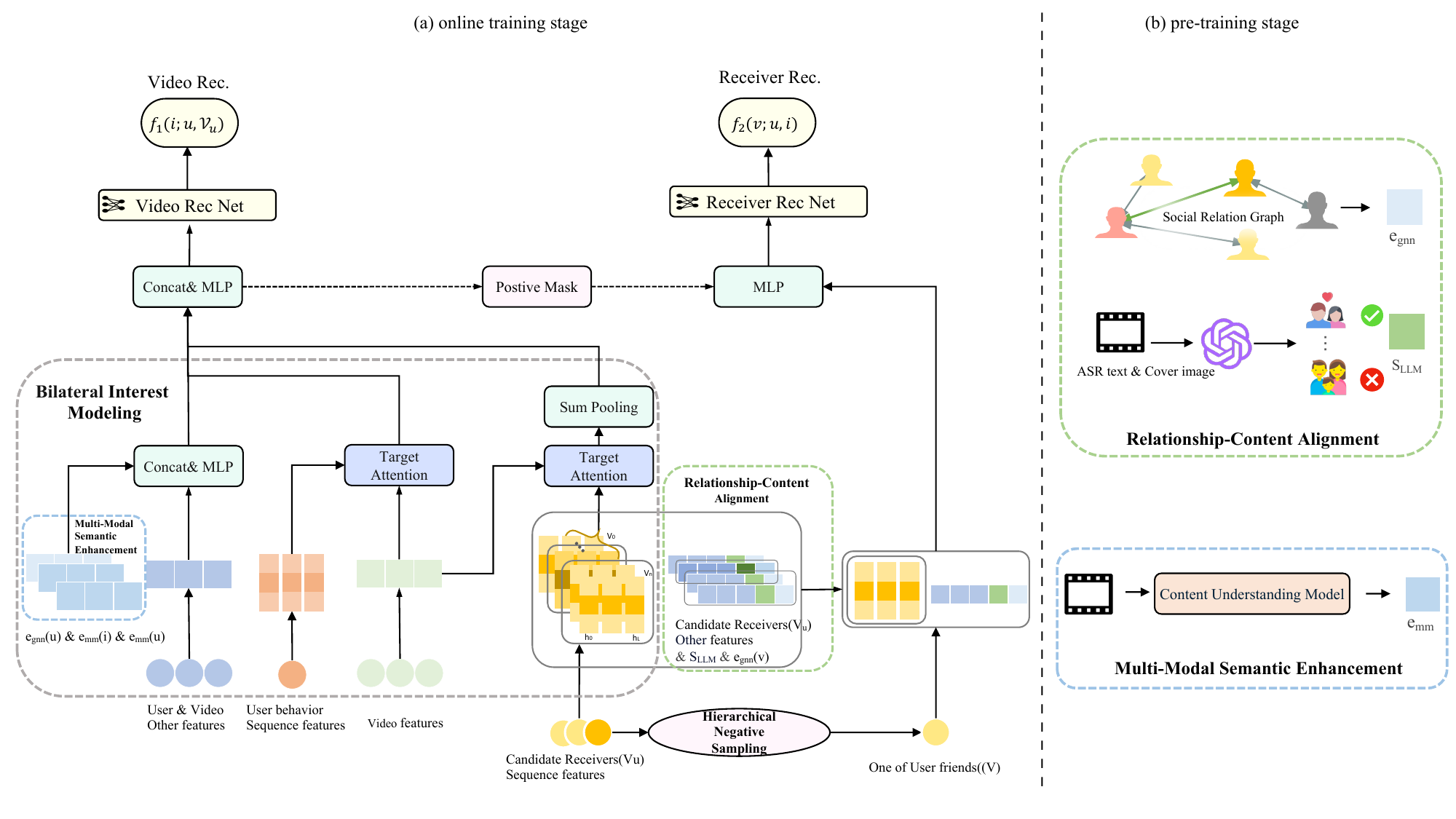}
  \caption{Overview of the UniShare Framework. UniShare jointly trains the video recommendation and receiver recommendation tasks, enhancing performance on both through underlying information sharing. Pre-computed GNN embeddings, multi-modal embeddings, and relationship alignment features are incorporated as enhanced inputs to the model.}
  \label{fig:unishare}
\end{figure*}
In this section, we present UniShare's joint modeling framework addressing the unique challenges of social sharing recommendation: bilateral satisfaction optimization and extreme data sparsity. Unlike traditional decoupled approaches, our method introduces dedicated modules to capture both the sharer-receiver-content relationship dynamics and alleviate sparse signal issues through multi-faceted enhancements. Overview of the framework is shown in figure \ref{fig:unishare}.

\subsection{Representation Enhancement for Bilateral Satisfaction}
Sharing behavior inherently embodies bilateral characteristics where both initiator satisfaction (sharer's intent) and receiver satisfaction (relevance and engagement) must be optimized simultaneously. To address this, we design enhanced representations through three dedicated components:

\subsubsection{\textbf{Bilateral Interest Modeling}}
We employ target attention mechanisms to dynamically model how specific content aligns with both parties' interests. The attention operation is formally defined as:
\begin{equation}
    \tilde{u} = \sum_{j=1}^{L}\alpha_j \cdot \mathbf{h}_j^u, \quad \alpha_j = \frac{\exp(\text{score}(i, \mathbf{h}_j^u))}{\sum_{k=1}^{L}\exp(\text{score}(i, \mathbf{h}_k^u))}
\end{equation}

where $\mathbf{h}_j^u$ represents the $j$-th historical interaction embedding of user $u$, and $\text{score}(\cdot)$ is implemented as a scaled dot-product attention function. Similarly, we compute receiver-side interest representation: $\tilde{v} = \text{TargetAttention}(i, \mathcal{H}_v)$. For the video recommendation task, the embedding representing the entire candidate receiver set $V_u$
is obtained by applying sum pooling to the embeddings of all individual receivers $v$

This architecture enables explicit modeling of whether content $i$ resonates with both $u$'s sharing preferences and $v$'s consumption preferences through learned attention weights that emphasize historically relevant interactions.

\paragraph{Multi-Modal Semantic Enhancement}Due to the extreme sparsity of sharing behavior, ID-based representations struggle to learn adequate embeddings for new or long-tail entities.To mitigate sparsity limitations of ID-based features, we incorporate pre-trained multi-modal embeddings:
(1) Video multi-modal embeddings $e_{mm}(i)$ from content understanding model. (2) User multi-modal interest embeddings $e_{mm}(u)$ aggregated from historical interactions. And then, we computing a matching score $s_{match} = e_{mm} (v) \cdot e_{mm}(i) $ between the candidate receeiver $v$ and the video $i$.

These continuous representations provide rich semantic signals that complement discrete ID features, particularly valuable for rare or new entities.
\subsubsection{\textbf{Relationship-Content Alignment}} We first introduce LLM-powered semantic matching between content characteristics and relationship types:  $s_{LLM} = \text{LLM}(\text{ASR}(i), cover(i),\text{relationship}_{uv})$. This module evaluates whether content $i$ is appropriate for the specific relationship type between $u$ and $v$ (e.g., sharing comedy clips with friends vs. educational content with parents). The prompt is provided in the appendix.

Furthermore, we incorporated pre-trained Graph Neural Network (GNN) embeddings, based on social relationships, to further enhance the relationship alignment. The graph for GNN training is constructed from the follow relationships within the recipient recommendation scenario. Edges between connected users form the positive samples. Global negative sampling is employed to generate negative samples. The model is trained with a contrastive loss objective to maximize the similarity between positive pairs and minimize the similarity between negative pairs.

Social graphs and content understanding serve as fundamental infrastructures within the Kuaishou platform. We directly provide those pre-trained embeddings in K-Share.

\subsection{Unified Learning Architecture for Joint Video and Receiver Recommendation }
We now introduce the methodology and training pipeline for the joint modeling of video and receiver recommendation. Kuaishou's baseline solution models and trains the share probability prediction tasks for video recommendation and receiver recommendation separately. 

This decoupled approach suffers from inherent limitations. The receiver recommendation task usually does not take the current video as input. Even if the video is used, the model suffers from the sparsity of sharing behavior: the number of shared videos is limited, leading to insufficient samples and inadequate learning of video features. Conversely, for the video recommendation task, learning in isolation struggles to effectively leverage features of the candidate receivers and their preferences, making it difficult to model bilateral satisfaction.

Our unified framework simultaneously addresses two tasks by following approachs.
\subsubsection{\textbf{Hierarchical Negative Sampling}}
To address the extreme sparsity of positive sharing signals, we implement a two-tier negative sampling strategy:  
\begin{itemize}
    \item Hard negatives: Users who have received shares from $u$ historically 
    \item Easy negatives: Random samples from $u$'s social connections
\end{itemize}
This design effectively addresses the extreme sparsity of positive signals in sharing behavior by adopting a two-stage negative sampling strategy—incorporating both hard negatives (historically shared recipients) and easy negatives (random social connections)—while simulating real-world interface which exposes a maximum of 20 candidate receivers, constraints to enhance model performance.

\subsubsection{\textbf{Parameter Sharing Strategy}}
The model shares embedding layers and feature transformation networks across both tasks, allowing knowledge transfer while maintaining task-specific heads:  
\begin{itemize}
    \item Video task: Uses full feature set including $u$-$i$ interaction features 
    \item Receiver task: Freezes $u$-$i$ parameters during backpropagation to maintain feature stability
\end{itemize}
It enables positive knowledge transfer to alleviate sparsity, ensures training stability through selective freezing, and maintains dedicated pathways for task-specific optimization, all while improving computational efficiency. This careful balance is a key reason for the framework's superior performance over decoupled baselines.

\subsubsection{\textbf{Multi-Task Optimization}}
We employ a joint training paradigm with weighted losses:  
\begin{equation}
\begin{split}
\mathcal{L}_{video} &= -\frac{1}{N} \sum_{(u, i)} [ y^{\text{s}}  \cdot \log(f_1(i; u, \mathcal{V}_u))  
\\ & + (1 - y^{\text{s}}) \cdot \log(1 - f_1(i; u, \mathcal{V}_u)) ]
\end{split}
\end{equation}
\begin{equation}
\mathcal{L}_{receiver} = -\frac{1}{M} \sum_{(u, i, v)} \left[ \log(f_2(v; u, i)) + \sum_{v^{\text{neg}}} \log(1 - f_2(v^{\text{neg}}; u, i)) \right]
\end{equation}
\begin{equation}
    \mathcal{L}_{total} = \mathcal{L}_{video} + \alpha\mathcal{L}_{receiver}
\end{equation}
where the video recommendation loss $\mathcal{L}_{video}$ leverages all exposure samples, while the receiver recommendation loss $\mathcal{L}_{receiver}$ utilizes only samples with sharing label, with corresponding nagetive samples curated from nagetive sampling strategy.

This architecture enables mutual enhancement between the two tasks: the video recommendation task benefits from receiver-side signals, while the receiver recommendation task leverages richer content understanding from the video modeling component.

\section{Experiment}

\subsection{Experiment Setup}
\subsubsection{Dataset}
\begin{table}
  \caption{Dataset Statistics.}
  \label{tab:dataset}
  \begin{tabular}{cccccc}
    \toprule
    \#Users & \#Videos & \#VideoViews & \#Shares & \#Friends & \#Receivers \\
    \midrule
    19395 & 3790673 & 11310319 & 752887 & 1306752 & 186886 \\
  \bottomrule
\end{tabular}
\end{table}
Video sharing represents an interactive behavior among users, content, and receivers. Specifically, on mainstream short-video platforms, we generally decompose this process into two tasks: video recommendation and receiver recommendation. We perform joint learning on these two tasks in this work. Currently, there is no publicly available dataset for this purpose. Therefore, we constructed \textit{K-Share}, a unified sharing modeling dataset, based on massive user and interaction logs from Kuaishou. K-Share comprises real data from a full month on the Kuaishou platform, including approximately 20,000 typical sharing users (with higher activity levels and numerous sharing receivers), encompassing about 1 million sharing instances and 10 million video view events.

Users were filtered to include those who were highly active in sharing and had between 1 and 10 distinct sharing receivers. Stratified sampling was applied based on the number of receivers, with sampling ratios of 1:3:5 for the strata of 1 receiver, 2-5 receivers, and over 5 receivers, respectively.

Requests were filtered to include only those with more than 3 
video impressions. Due to the sparsity of sharing behavior, while all requests containing positive sharing samples were retained, only 5\% of requests without positive samples were randomly sampled to manage the immense data volume.

At the feature level, the dataset provides: (1) Basic user attributes. (2) Historical sequences of sharing initiation and receipt. (3) Pre-trained multi-modal interest embeddings, GNN embedding based on user relations. (3) Basic video information. (4) Pre-trained multi-modal video embeddings.

For the receiver recommendation task and the joint modeling framework, the dataset preserves: (1) The receiver ID for positive sharing instances. (2) The complete friend list for each user. (3) The list of users with whom sharing occurred in the past year. (4) For each user, a set of up to 20 candidate receivers was generated for evaluation purposes.

The basic statistics of the dataset are presented in the Table \ref{tab:dataset}. We split the data into training, validation, and test sets with an 8:1:1 ratio according to the timestamp of share action order. For the receiver recommendation task, we generate candidate receivers for evaluation using the same negative sampling strategy used during training. Note that the pool of candidate users for a given user may differ between the training and evaluation.

\subsubsection{Baseline Models}
To the best of our knowledge, no existing work has yet explored a joint modeling approach for this problem. Therefore, we select separate models for video recommendation and receiver recommendation as our baselines. Specifically, the video recommendation baseline employs a multi-task PLE (Progressive Layered Extraction)\cite{PLE} model, while the receiver recommendation baseline utilizes a DCN (Deep \& Cross Network)\cite{DCNv2} model.
For the experimental group employing joint modeling, we adopt an identical top-level model architecture. Furthermore, we introduce several ablation groups to rigorously validate the contribution of both the joint modeling framework and the proposed representation enhancements. These ablation groups include:(1)w/o all: Joint training only. (2)w/o RCA: Removes the relationship-content alignment( including the GNN embeddings). (3)w/o BIM: Removes the bilateral interest modeling (4)w/o HNS: Removes the hierarchical negative sampling strategy.

\subsubsection{Evaluation Metrics}
The two sub-tasks of UniShare are applied to video ranking and receiver ranking, respectively. Therefore, we primarily employ \textbf{AUC}, \textbf{GAUC}, \textbf{Recall@K}, \textbf{NDCG@K}, and \textbf{MRR} to evaluate the model's ranking capability.
For the video recommendation task, the evaluation is performed on all actual impression samples without additional processing. Except for AUC, all metrics are computed by first calculating the metric per user and then averaging the results. We set $K=10$.
For the receiver recommendation task, the evaluation uses the actual impression data from the share panel interface. Negative samples are supplemented using the negative sampling strategy until a maximum of 20 candidates per impression is reached. All metrics are computed per request (i.e., per $\langle u, i \rangle$ pair) and then averaged.We set  $K=5$, as the first page of the share panel exposes exactly 5 candidate receivers.

\subsubsection{Implementation Details} 

For all models, we maintained consistent batch sizes and embedding dimensions. For the jointly trained model, aside from the differences inherent to joint training and enhanced representations, we kept the top-level model architecture identical to the baseline. We set the loss weights to  $\alpha$=2 for the video recommendation and receiver recommendation tasks, respectively. The Adam optimizer was employed, and the learning rate was tuned within the set \{0.005, 0.003, 0.001, 0.0005, 0.0001\}.The pre-trained embeddings provided in the K-Share dataset were processed through an identical DNN network to compress their dimensionality to 64. For the relationship matching module, the Qwen-VL\cite{Qwen-VL} model was utilized to generate the relationship-content alignment results.The relationship types include \textit{romantic partners}, \textit{ buddies }, \textit{sister-friend}, \textit{elders}, \textit{juniors}, and  \textit{peers}, covering approximately 80\% of $<u,v>$ pairs in K-Share.

\begin{table*}
\caption{Performance comparison and ablation study on K-Share. The best and best results are highlighted in bold. “*” denotes that the improvements are significant($p$-value < 0.05)}
\label{tab:results}
\begin{tabular}{lllllllllll}
\toprule
\multicolumn{1}{l}{\multirow{2}{*}{Model}} & \multicolumn{5}{l}{Video Rec}          & \multicolumn{5}{l}{Receiver Rec}       \\
\multicolumn{1}{l}{}                       & AUC & GAUC & NDCG@10 & RECALL@10 & MRR & AUC & GAUC & NDCG@5 & RECALL@5 & MRR \\
\midrule
\midrule
PLE (video)   &  0.7512&  0.6919&  0.3099&  0.2869&  0.3880&  -&  -&  -&  -&  -\\
DCN (receiver) &  -&  -&  -&  -&  -&  0.9124&  0.9140&  0.9046&  0.8428& 0.8456\\
\midrule
\textbf{UniShare}& \textbf{0.7588}*& \textbf{0.6976}*& \textbf{0.3110} & \textbf{0.2913}* & \textbf{0.3934}*& \textbf{0.9307}*& \textbf{0.9282}*& \textbf{0.9093}*& \textbf{0.8532}*&\textbf{0.8568}*\\
 w/o all & 0.7526& 0.6930& 0.3103& 0.2892& 0.3903& 0.9213& 0.9248& 0.8995& 0.8394&0.8422\\
 w/o RCA& 0.7566& 0.6946& 0.3106& 0.2901& 0.3918& 0.9282& 0.9267& 0.9073& 0.8511&0.8554\\
 w/o BIM& 0.7558& 0.6940& 0.3106& 0.2899& 0.3916& 0.9273& 0.9266& 0.9069& 0.8499&0.8538\\
 w/o HNS& 0.7584& 0.6975& 0.3110& 0.2911& 0.3932& 0.9238& 0.9259& 0.9051& 0.8468&0.8499\\
\bottomrule
\end{tabular}
\end{table*}

\subsection{Experiment Results}
The overall results of 2 tasks are shown in Table \ref{tab:results}. Comprehensive experimental results on the K-Share dataset demonstrate that UniShare achieves significant improvements over strong baselines across both recommendation tasks. The unified framework substantially outperforms separately trained models, validating the effectiveness of joint modeling for ternary interaction recommendation.

\subsubsection{Video Recommendation Experiment Results}
UniShare demonstrates remarkable improvements in video recommendation performance, achieving an AUC of 0.7588 compared to the PLE baseline's 0.7512, representing a statistically significant improvement of +1.01\%. The GAUC metric shows a similar trend, with UniShare reaching 0.6976 versus the baseline's 0.6919 (+0.82\% improvement). More importantly, the Recall@10 metric shows a substantial +1.53\% improvement (0.2913 vs. 0.2869), indicating that UniShare successfully identifies more share-worthy videos in the top recommendations. These results collectively demonstrate that joint modeling with receiver recommendation signals significantly enhances video sharing prediction accuracy, particularly in the critical top-ranked items that directly impact user experience.

\subsubsection{Receiver Recommendation Task}
The receiver recommendation task shows even more substantial gains, with UniShare achieving an AUC of 0.9307 compared to the DCN baseline's 0.9124, representing a significant +2.01\% improvement. The GAUC metric follows this trend with a +1.55\% improvement (0.9282 vs. 0.9140). The practical impact is evident in the Recall@5 metric, which shows a +1.23\% improvement (0.8532 vs. 0.8428), demonstrating that UniShare more effectively identifies the most appropriate receivers in the limited visible candidate set. The MRR improvement of +1.32\% further confirms that the correct receivers are ranked higher in the recommendation list. These results underscore the critical importance of incorporating video content information and joint training for accurate receiver recommendation.

\subsubsection{Ablation Study Analysis} 
The ablation analysis reveals the individual contributions of each component in the UniShare framework. The complete removal of enhanced representations ("w/o all") causes the most substantial performance degradation, with video AUC dropping to 0.7526 (-0.82\% from full UniShare) and receiver AUC decreasing to 0.9213 (-1.01\%), highlighting the fundamental importance of the proposed representation enhancements. The relationship-content alignment module (RCA) contributes significantly, as its removal results in a 0.29\% drop in video AUC and a 0.27\% drop in receiver AUC. Similarly, the bilateral interest modeling (BIM) component shows its value through a 0.40\% video AUC decrease and 0.37\% receiver AUC decrease when omitted. Most notably, the hierarchical negative sampling strategy (HNS) proves particularly crucial for receiver recommendation, with its removal causing a 0.74\% drop in receiver AUC while leaving video metrics largely unchanged, demonstrating its specialized role in addressing the unique challenges of receiver ranking.

The collective findings from these ablation experiments confirm that each proposed component—enhanced representations, relationship-content alignment, bilateral interest modeling, and hierarchical negative sampling—makes distinct and valuable contributions to the overall performance of the unified framework.

\subsubsection{Satisfaction of Receivers} Modeling sharing behavior requires the simultaneous optimization of satisfaction for both the User (sharer) and the Receiver. In online A/B testing, we can evaluate Receiver satisfaction by observing metrics such as their view-through rate and reply rate on shared content. However, this direct evaluation is not feasible on offline datasets.Therefore, we conducted an offline analysis by calculating a matching score between the interest embeddings of the historical receivers and the top 10 videos with the highest share probability for each user in the test set. This analysis revealed an average increase in the matching score of +3.46\%. A detailed evaluation of receiver satisfaction will be comprehensively carried out in our online A/B tests.

\begin{figure}[h]
  \centering
  \includegraphics[width=0.8\linewidth]{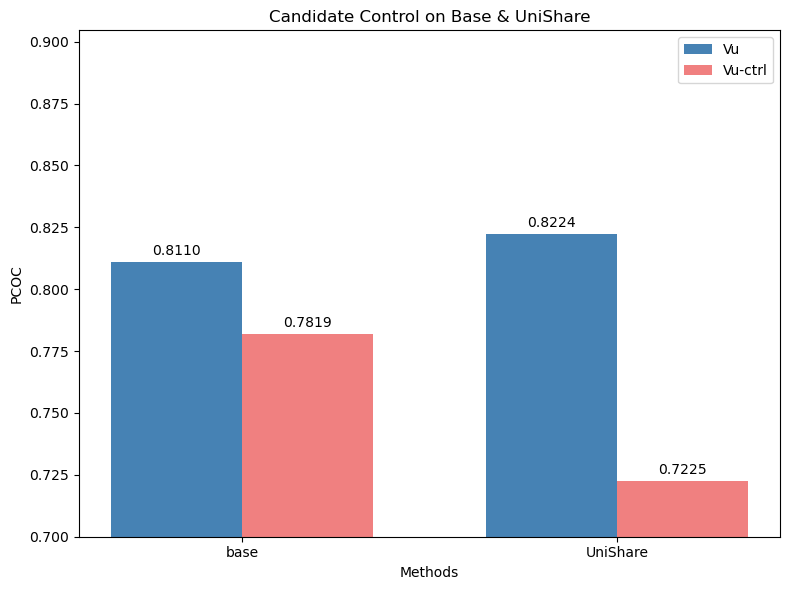}
  \caption{Comparison of PCOC change when modifying the input candidate set  $\mathcal{V}_u$. $\mathcal{V}_u$-ctrl denotes the removal of the most frequently shared receiver from the candidate set.}
  \label{fig:Vu-ctrl}
\end{figure}
We experimented with modifying the candidate receiver set $\mathcal{V}_u$ during inference by specifically removing the most frequently shared receiver. We then observed the change in the estimated PCOC (predicted click over click) for the video recommendation task in both the base model and the UniShare model.As shown in figure \ref{fig:Vu-ctrl}, the change in PCOC for the base model was significantly smaller than that for the UniShare model. This indicates that, through unified modeling and enhanced representation, UniShare is better able to capture and utilize information from the candidate receiver set.

\begin{figure}[h]
  \centering
  \includegraphics[width=0.8\linewidth]{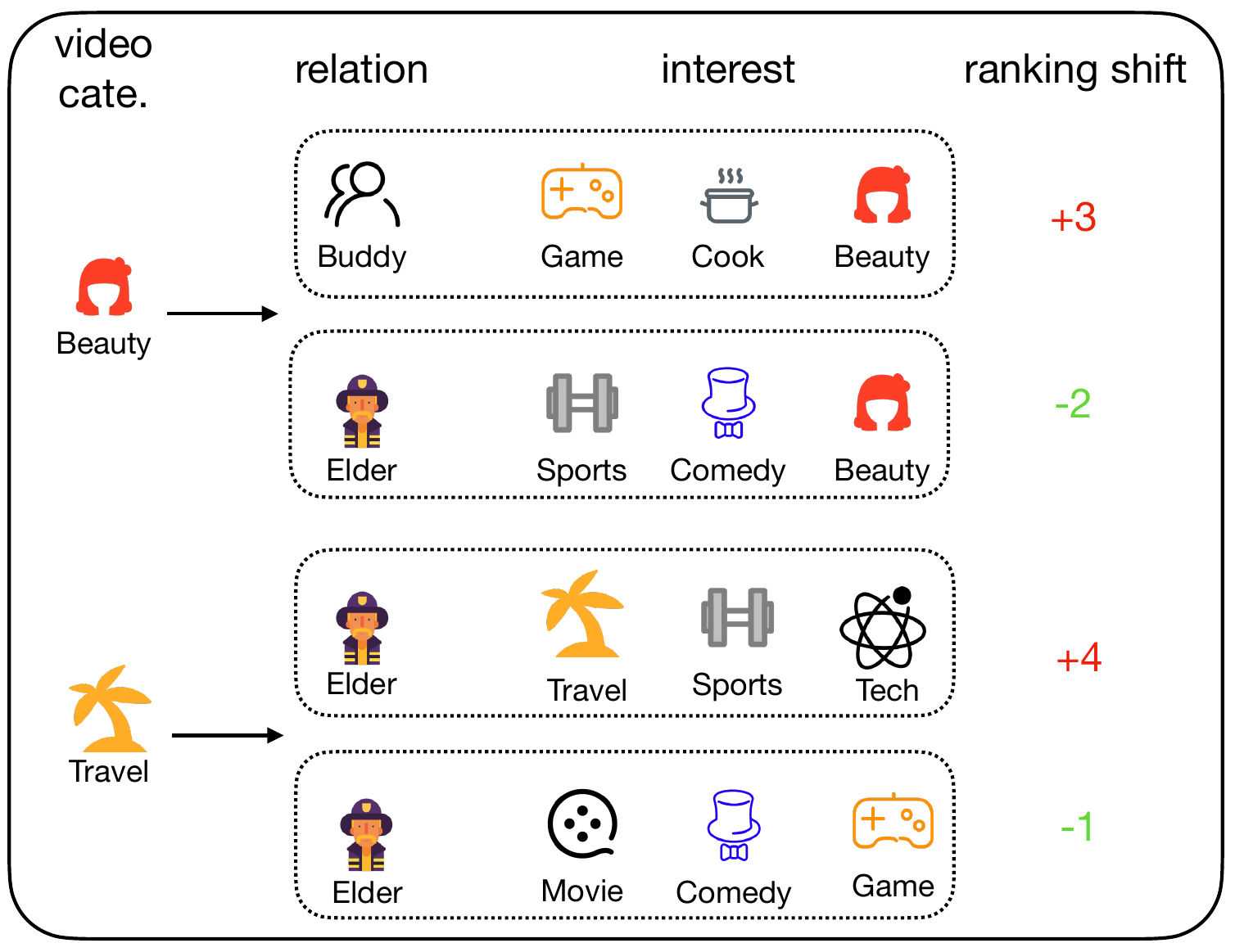}
  \caption{Example illustration of the effects of RCA and BIM}
  \label{fig:case}
\end{figure}

\paragraph{Case Study}
Finally, we conduct case studies to demonstrate the effectiveness of UniShare in receiver-side matching and satisfaction enhancement. As illustrated in Figure \ref{fig:case}, we selected two distinct users with representative video sharing instances, along with two candidate receivers per case. For each case, we compare the ranking differences between the baseline  and the UniShare results. Case1 show that both candidate users exhibit similar interests in the Beauty. However, the video content is more suitable for sharing with a Buddy rather than an Older. The candidate with the Buddy relationship ascended by 3 positions in UniShare’s ranking, demonstrating that RCA effectively aligns relationship compatibility with content type.
Case2 show that Both candidate receivers share an Older relationship with the sharer, but their interest preferences differ significantly. UniShare’s ranking elevated the candidate with a preference for travel-related videos by 4 positions, highlighting the contribution of Bilateral Interest Modeling (BIM) in capturing nuanced preference matching beyond relational constraints.

These cases confirm that UniShare’s integrated RCA and BIM mechanisms synergistically enhance receiver selection by jointly optimizing relationship appropriateness and interest relevance.

\subsection{Online Deployment}
\label{sec:online-ab}
\begin{figure}[h]
  \centering
  \includegraphics[width=0.9\linewidth]{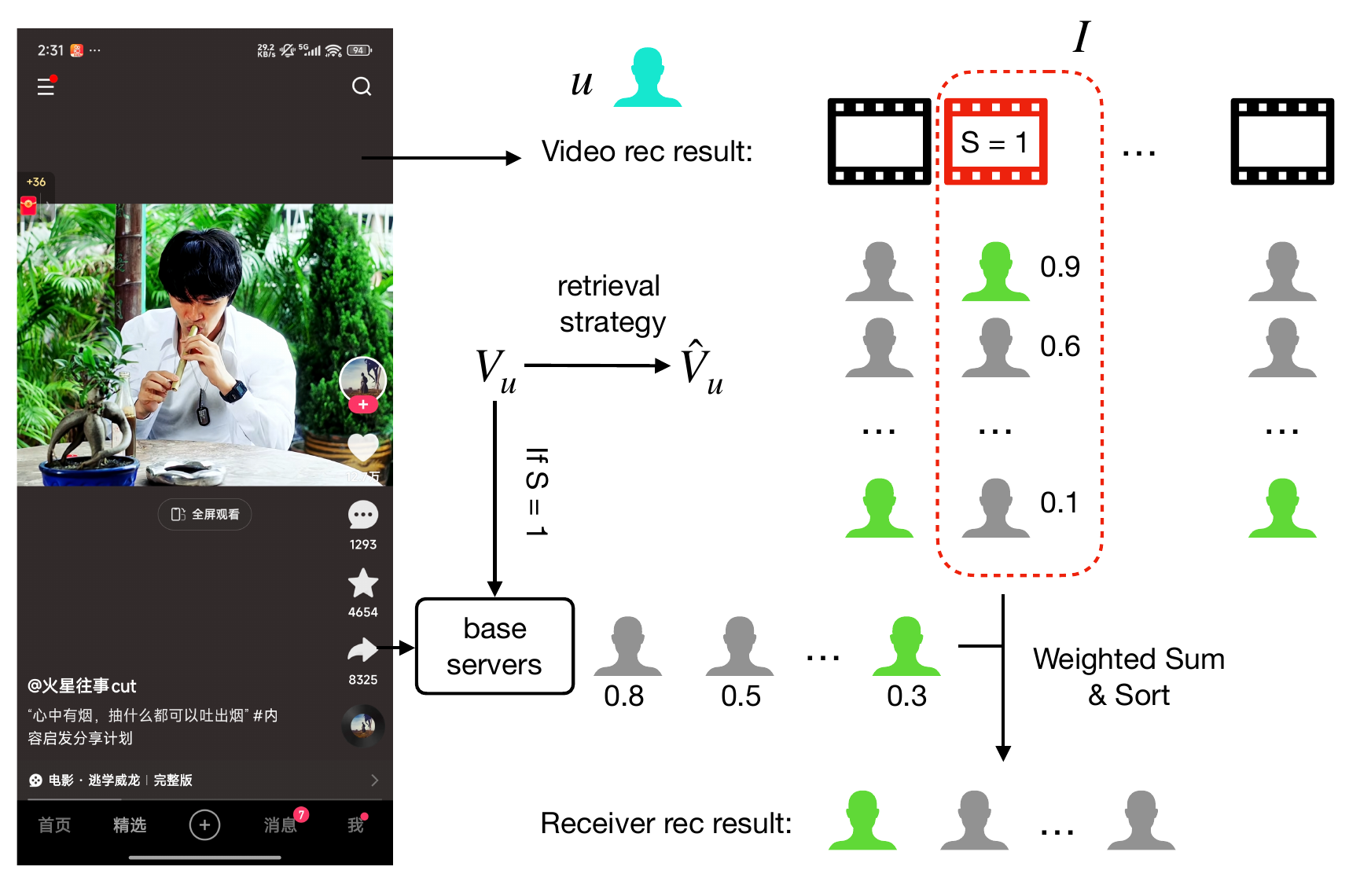}
  \caption{For a single user request, a joint recommendation approach requires estimating a subsequent receivers ranking for every candidate video. We retrieve a subset $\mathcal{\hat{V}}_u$ from $\mathcal{V}_u$ for UniShare. The final output is produced by merging and re-ranking the results from this subset with those from the full candidate set.}
  \label{fig:deployment}
\end{figure}
\subsubsection{Deployment and Application at KuaiShou}
Deploying the joint video-receiver recommendation system online introduces a significant challenge of escalating computational complexity. The existing solution deploys video recommendation and receiver recommendation as separate services. When a user consumes videos, the video recommendation service is invoked. It estimates signals such as the share probability, followed by a ranking stage to produce the final feed. The number of samples scored per request in this service is proportional to the size of the video candidate pool, $|\mathcal{I}|$. Subsequently, only when a user clicks the share button is the receiver recommendation service invoked, which ranks the candidate receivers. The number of samples scored per request here is proportional to the size of the user candidate pool for that user, $|\mathcal{V}_u|$. Due to the sparsity of sharing actions, the Queries Per Second (QPS) for the receiver service is considerably lower than that for the video service.

In contrast, a joint recommendation approach requires, for a single user request, estimating a subsequent receiver ranking \textbf{for every candidate video}. This results in a total sample scoring complexity of $|\mathcal{I}| \times |\mathcal{V}_u|$ per request. Furthermore, the invocation logic changes: scoring for the receiver task must now occur even if the share button is never clicked.

To address this, We have designed an online deployment scheme as illustrated in Figure \ref{fig:deployment}. We employ a simple retrieval strategy to drastically reduce the effective size of $\mathcal{V}_u$ during deployment. For the online experimental version, we directly retrieve up to the 8 most recently shared-with users. We maintain the original system's architecture that performs scoring over the full candidate sets. The scores are then used as a key feature for a final reranking stage that produces the ultimate output. Specifically, for candidates covered by the UniShare model, their final score is calculated as the weighted sum of the UniShare score and the base model score. For candidates not covered by the UniShare model, the final score is derived from the weighted sum of two base model scores.This strategy offers significant operational flexibility. The retrieval policy for $\mathcal{V}_u$ can be strategically aligned with business objectives. For instance, based on analysis suggesting that shares to low-activity users yield higher platform value, the retrieval strategy can be biased to prioritize such users. This policy simultaneously influences both sub-tasks: if a user $u$ has no low-activity candidates in their retrieved set ($\mathcal{V}_u = \emptyset$), the estimated overall share probability $f_1(i; u, \mathcal{V}_u)$ for any video $i$ will be small. Consequently, in the subsequent multi-task ranking stage, the model will prioritize other engagement objectives over sharing for this user.

\subsubsection{Online A/B Testing Results}
Following the online deployment of the proposed solution, we conducted a rigorous online A/B experiment over a period of 7 days. We primarily evaluated 5 key metrics: the number of shares, the number of unique sharing users, the click-through rate (CTR) of the share button, the CTR of icon on the sharing panel and reply rate of receivers.
The number of shares and the number of unique sharing users are utilized to assess the overall performance of sharing behavior. The CTR of the share button serves to evaluate the effectiveness of the video recommendation sub-task, while the CTR of the sharing panel evaluates the performance of the receiver recommendation sub-task.The reply rate of receivers serves to evaluate the satisfaction of receivers.

The final results in Table \ref{tab:online-ab} demonstrate that UniShare achieved statistically significant improvements across all 5 metrics. The specific results are as follows: (1) Receiver Recommendation Task Performance: CTR of the share icon on the sharing panel: +1.14\%. (2) Video Recommendation Task Performance: CTR of the share button: +1.12\%. (3) Overall Sharing Engagement: Number of shares: +1.95\% \& Number of unique sharing users: +0.805\%. (4) Receiver Satisfaction Metric: Reply rate of receivers: +0.482\%.

\begin{table}
  \caption{Online A/B Testing Results on KuaiShou platform}
  \label{tab:online-ab}
  \begin{tabular}{cc}
    \toprule
    \textbf{Online Metrics} & \textbf{Improvement} \\
    \midrule
    
    number of shares & +1.95\% \\
    number of unique sharing users & +0.805\%  \\
    CTR of the share button & +1.12\%  \\
    CTR of icon on the sharing panel & +1.14\% \\
    reply rate of receivers & +0.482\% \\
  \bottomrule
\end{tabular}
\end{table}

\section{Conclusion}
In this paper, we identified the critical problem of disjointed modeling for video sharing and receiver recommendation in social content platforms. We proposed \textbf{UniShare}, a unified framework that jointly learns to predict the sharing probability.
The core of UniShare lies in its enhanced representation learning, which leverages pre-trained social and multi-modal embeddings, and its explicit modeling of bilateral interest and relationship-content alignment. More importantly, the joint training paradigm allows the two tasks to inform and enhance each other, effectively alleviating the data sparsity issue inherent in sharing behavior and leading to a more accurate estimation of bilateral satisfaction.
Extensive evaluations on the curated \textbf{K-Share} dataset demonstrated UniShare's superior performance over separate state-of-the-art baselines on both recommendation tasks. The successful online deployment and significant improvements observed in the A/B test on the Kuaishou platform further validate its practical value and effectiveness in a real-world production environment.

For future work, we plan to explore more efficient inference architectures to reduce the computational overhead of joint scoring. Additionally, investigating more advanced generative recommendation are promising directions to further enhance the framework.

\bibliographystyle{ACM-Reference-Format}
\bibliography{sample-base}

@String{Computer = "{IEEE} Computer" }

@String{Chelsea = "Chelsea" }

@inproceedings{HGSRec,
  author       = {Houye Ji and
                  Junxiong Zhu and
                  Xiao Wang and
                  Chuan Shi and
                  Bai Wang and
                  Xiaoye Tan and
                  Yanghua Li and
                  Shaojian He},
  title        = {Who You Would Like to Share With? {A} Study of Share Recommendation
                  in Social E-commerce},
  booktitle    = {Thirty-Fifth {AAAI} Conference on Artificial Intelligence, {AAAI}
                  2021, Thirty-Third Conference on Innovative Applications of Artificial
                  Intelligence, {IAAI} 2021, The Eleventh Symposium on Educational Advances
                  in Artificial Intelligence, {EAAI} 2021, Virtual Event, February 2-9,
                  2021},
  pages        = {232--239},
  publisher    = {{AAAI} Press},
  year         = {2021},
  url          = {https://doi.org/10.1609/aaai.v35i1.16097},
  doi          = {10.1609/AAAI.V35I1.16097},
  bibsource    = {dblp computer science bibliography, https://dblp.org}
}

@inproceedings{ReSeq,
  author       = {Bowen Zheng and
                  Yupeng Hou and
                  Wayne Xin Zhao and
                  Yang Song and
                  Hengshu Zhu},
  editor       = {Jie Zhang and
                  Li Chen and
                  Shlomo Berkovsky and
                  Min Zhang and
                  Tommaso Di Noia and
                  Justin Basilico and
                  Luiz Pizzato and
                  Yang Song},
  title        = {Reciprocal Sequential Recommendation},
  booktitle    = {Proceedings of the 17th {ACM} Conference on Recommender Systems, RecSys
                  2023, Singapore, Singapore, September 18-22, 2023},
  pages        = {89--100},
  publisher    = {{ACM}},
  year         = {2023},
  url          = {https://doi.org/10.1145/3604915.3608798},
  doi          = {10.1145/3604915.3608798},
  bibsource    = {dblp computer science bibliography, https://dblp.org}
}

@inproceedings{DynShare,
  author       = {Ziwei Zhao and
                  Xi Zhu and
                  Tong Xu and
                  Aakas Lizhiyu and
                  Yu Yu and
                  Xueying Li and
                  Zikai Yin and
                  Enhong Chen},
  editor       = {Hsin{-}Hsi Chen and
                  Wei{-}Jou (Edward) Duh and
                  Hen{-}Hsen Huang and
                  Makoto P. Kato and
                  Josiane Mothe and
                  Barbara Poblete},
  title        = {Time-interval Aware Share Recommendation via Bi-directional Continuous
                  Time Dynamic Graphs},
  booktitle    = {Proceedings of the 46th International {ACM} {SIGIR} Conference on
                  Research and Development in Information Retrieval, {SIGIR} 2023, Taipei,
                  Taiwan, July 23-27, 2023},
  pages        = {822--831},
  publisher    = {{ACM}},
  year         = {2023},
  url          = {https://doi.org/10.1145/3539618.3591775},
  doi          = {10.1145/3539618.3591775},
  bibsource    = {dblp computer science bibliography, https://dblp.org}
}

@inproceedings{DPGNN,
  author       = {Chen Yang and
                  Yupeng Hou and
                  Yang Song and
                  Tao Zhang and
                  Ji{-}Rong Wen and
                  Wayne Xin Zhao},
  editor       = {Jennifer Golbeck and
                  F. Maxwell Harper and
                  Vanessa Murdock and
                  Michael D. Ekstrand and
                  Bracha Shapira and
                  Justin Basilico and
                  Keld T. Lundgaard and
                  Even Oldridge},
  title        = {Modeling Two-Way Selection Preference for Person-Job Fit},
  booktitle    = {RecSys '22: Sixteenth {ACM} Conference on Recommender Systems, Seattle,
                  WA, USA, September 18 - 23, 2022},
  pages        = {102--112},
  publisher    = {{ACM}},
  year         = {2022},
  url          = {https://doi.org/10.1145/3523227.3546752},
  doi          = {10.1145/3523227.3546752},
  bibsource    = {dblp computer science bibliography, https://dblp.org}
}

@article{SocialLGN,
  author       = {Jie Liao and
                  Wei Zhou and
                  Fengji Luo and
                  Junhao Wen and
                  Min Gao and
                  Xiuhua Li and
                  Jun Zeng},
  title        = {SocialLGN: Light graph convolution network for social recommendation},
  journal      = {Inf. Sci.},
  volume       = {589},
  pages        = {595--607},
  year         = {2022},
  url          = {https://doi.org/10.1016/j.ins.2022.01.001},
  doi          = {10.1016/J.INS.2022.01.001},
  bibsource    = {dblp computer science bibliography, https://dblp.org}
}

@article{Social-RippleNet,
  author       = {Wenbo Jiang and
                  Yanrui Sun},
  title        = {Social-RippleNet: Jointly modeling of ripple net and social information
                  for recommendation},
  journal      = {Appl. Intell.},
  volume       = {53},
  number       = {3},
  pages        = {3472--3487},
  year         = {2023},
  url          = {https://doi.org/10.1007/s10489-022-03620-2},
  doi          = {10.1007/S10489-022-03620-2},
  bibsource    = {dblp computer science bibliography, https://dblp.org}
}

@inproceedings{DGNN,
  author       = {Lianghao Xia and
                  Yizhen Shao and
                  Chao Huang and
                  Yong Xu and
                  Huance Xu and
                  Jian Pei},
  title        = {Disentangled Graph Social Recommendation},
  booktitle    = {39th {IEEE} International Conference on Data Engineering, {ICDE} 2023,
                  Anaheim, CA, USA, April 3-7, 2023},
  pages        = {2332--2344},
  publisher    = {{IEEE}},
  year         = {2023},
  url          = {https://doi.org/10.1109/ICDE55515.2023.00180},
  doi          = {10.1109/ICDE55515.2023.00180},
  bibsource    = {dblp computer science bibliography, https://dblp.org}
}

@article{Qwen-VL,
  title={Qwen-VL: A Frontier Large Vision-Language Model with Versatile Abilities},
  author={Bai, Jinze and Bai, Shuai and Yang, Shusheng and Wang, Shijie and Tan, Sinan and Wang, Peng and Lin, Junyang and Zhou, Chang and Zhou, Jingren},
  journal={arXiv preprint arXiv:2308.12966},
  year={2023}
}

@inproceedings{DCNv2,
  author       = {Ruoxi Wang and
                  Rakesh Shivanna and
                  Derek Zhiyuan Cheng and
                  Sagar Jain and
                  Dong Lin and
                  Lichan Hong and
                  Ed H. Chi},
  editor       = {Jure Leskovec and
                  Marko Grobelnik and
                  Marc Najork and
                  Jie Tang and
                  Leila Zia},
  title        = {{DCN} {V2:} Improved Deep {\&} Cross Network and Practical Lessons
                  for Web-scale Learning to Rank Systems},
  booktitle    = {{WWW} '21: The Web Conference 2021, Virtual Event / Ljubljana, Slovenia,
                  April 19-23, 2021},
  pages        = {1785--1797},
  publisher    = {{ACM} / {IW3C2}},
  year         = {2021},
  url          = {https://doi.org/10.1145/3442381.3450078},
  doi          = {10.1145/3442381.3450078},
  bibsource    = {dblp computer science bibliography, https://dblp.org}
}

@inproceedings{PLE,
  author       = {Hongyan Tang and
                  Junning Liu and
                  Ming Zhao and
                  Xudong Gong},
  editor       = {Rodrygo L. T. Santos and
                  Leandro Balby Marinho and
                  Elizabeth M. Daly and
                  Li Chen and
                  Kim Falk and
                  Noam Koenigstein and
                  Edleno Silva de Moura},
  title        = {Progressive Layered Extraction {(PLE):} {A} Novel Multi-Task Learning
                  {(MTL)} Model for Personalized Recommendations},
  booktitle    = {RecSys 2020: Fourteenth {ACM} Conference on Recommender Systems, Virtual
                  Event, Brazil, September 22-26, 2020},
  pages        = {269--278},
  publisher    = {{ACM}},
  year         = {2020},
  url          = {https://doi.org/10.1145/3383313.3412236},
  doi          = {10.1145/3383313.3412236},
  bibsource    = {dblp computer science bibliography, https://dblp.org}
}

@article{Joint-Model-Survay,
  author       = {Xiangyu Zhao and
                  Yichao Wang and
                  Bo Chen and
                  Jingtong Gao and
                  Yuhao Wang and
                  Xiaopeng Li and
                  Pengyue Jia and
                  Qidong Liu and
                  Huifeng Guo and
                  Ruiming Tang},
  title        = {Joint Modeling in Recommendations: {A} Survey},
  journal      = {CoRR},
  volume       = {abs/2502.21195},
  year         = {2025},
  url          = {https://doi.org/10.48550/arXiv.2502.21195},
  doi          = {10.48550/ARXIV.2502.21195},
  eprinttype    = {arXiv},
  eprint       = {2502.21195},
  bibsource    = {dblp computer science bibliography, https://dblp.org}
}

@inproceedings{STAR,
  author       = {Xiang{-}Rong Sheng and
                  Liqin Zhao and
                  Guorui Zhou and
                  Xinyao Ding and
                  Binding Dai and
                  Qiang Luo and
                  Siran Yang and
                  Jingshan Lv and
                  Chi Zhang and
                  Hongbo Deng and
                  Xiaoqiang Zhu},
  editor       = {Gianluca Demartini and
                  Guido Zuccon and
                  J. Shane Culpepper and
                  Zi Huang and
                  Hanghang Tong},
  title        = {One Model to Serve All: Star Topology Adaptive Recommender for Multi-Domain
                  {CTR} Prediction},
  booktitle    = {{CIKM} '21: The 30th {ACM} International Conference on Information
                  and Knowledge Management, Virtual Event, Queensland, Australia, November
                  1 - 5, 2021},
  pages        = {4104--4113},
  publisher    = {{ACM}},
  year         = {2021},
  url          = {https://doi.org/10.1145/3459637.3481941},
  doi          = {10.1145/3459637.3481941},
  bibsource    = {dblp computer science bibliography, https://dblp.org}
}

@inproceedings{ESMM,
  author       = {Xiao Ma and
                  Liqin Zhao and
                  Guan Huang and
                  Zhi Wang and
                  Zelin Hu and
                  Xiaoqiang Zhu and
                  Kun Gai},
  editor       = {Kevyn Collins{-}Thompson and
                  Qiaozhu Mei and
                  Brian D. Davison and
                  Yiqun Liu and
                  Emine Yilmaz},
  title        = {Entire Space Multi-Task Model: An Effective Approach for Estimating
                  Post-Click Conversion Rate},
  booktitle    = {The 41st International {ACM} {SIGIR} Conference on Research {\&}
                  Development in Information Retrieval, {SIGIR} 2018, Ann Arbor, MI,
                  USA, July 08-12, 2018},
  pages        = {1137--1140},
  publisher    = {{ACM}},
  year         = {2018},
  url          = {https://doi.org/10.1145/3209978.3210104},
  doi          = {10.1145/3209978.3210104},
  bibsource    = {dblp computer science bibliography, https://dblp.org}
}

@inproceedings{MMOE,
  author       = {Jiaqi Ma and
                  Zhe Zhao and
                  Xinyang Yi and
                  Jilin Chen and
                  Lichan Hong and
                  Ed H. Chi},
  editor       = {Yike Guo and
                  Faisal Farooq},
  title        = {Modeling Task Relationships in Multi-task Learning with Multi-gate
                  Mixture-of-Experts},
  booktitle    = {Proceedings of the 24th {ACM} {SIGKDD} International Conference on
                  Knowledge Discovery {\&} Data Mining, {KDD} 2018, London, UK,
                  August 19-23, 2018},
  pages        = {1930--1939},
  publisher    = {{ACM}},
  year         = {2018},
  url          = {https://doi.org/10.1145/3219819.3220007},
  doi          = {10.1145/3219819.3220007},
  bibsource    = {dblp computer science bibliography, https://dblp.org}
}

@inproceedings{MBGCN,
  title={Multi-behavior recommendation with graph convolutional networks},
  author={Jin, Bowen and Gao, Chen and He, Xiangnan and Jin, Depeng and Li, Yong},
  booktitle={Proceedings of the 43rd international ACM SIGIR conference on research and development in information retrieval},
  pages={659--668},
  year={2020}
}

@inproceedings{SNR,
  title={Snr: Sub-network routing for flexible parameter sharing in multi-task learning},
  author={Ma, Jiaqi and Zhao, Zhe and Chen, Jilin and Li, Ang and Hong, Lichan and Chi, Ed H},
  booktitle={Proceedings of the AAAI conference on artificial intelligence},
  volume={33},
  number={01},
  pages={216--223},
  year={2019}
}

@article{LHUC,
  title={Learning hidden unit contributions for unsupervised acoustic model adaptation},
  author={Swietojanski, Pawel and Li, Jinyu and Renals, Steve},
  journal={IEEE/ACM Transactions on Audio, Speech, and Language Processing},
  volume={24},
  number={8},
  pages={1450--1463},
  year={2016},
  publisher={IEEE}
}

@inproceedings{ESM2,
  title={Entire space multi-task modeling via post-click behavior decomposition for conversion rate prediction},
  author={Wen, Hong and Zhang, Jing and Wang, Yuan and Lv, Fuyu and Bao, Wentian and Lin, Quan and Yang, Keping},
  booktitle={Proceedings of the 43rd International ACM SIGIR conference on research and development in Information Retrieval},
  pages={2377--2386},
  year={2020}
}

@misc{DBMTL,
  title={Deep Bayesian multi-target learning for recommender systems. CoRR abs/1902.09154 (2019)},
  author={Wang, Qi and Ji, Zhihui and Liu, Huasheng and Zhao, Binqiang},
  year={1902}
}

@inproceedings{EHCF,
  title={Efficient heterogeneous collaborative filtering without negative sampling for recommendation},
  author={Chen, Chong and Zhang, Min and Zhang, Yongfeng and Ma, Weizhi and Liu, Yiqun and Ma, Shaoping},
  booktitle={Proceedings of the AAAI conference on artificial intelligence},
  volume={34},
  number={01},
  pages={19--26},
  year={2020}
}

@article{GMSL,
  title={Generator and critic: A deep reinforcement learning approach for slate re-ranking in e-commerce},
  author={Wei, Jianxiong and Zeng, Anxiang and Wu, Yueqiu and Guo, Peng and Hua, Qingsong and Cai, Qingpeng},
  journal={arXiv preprint arXiv:2005.12206},
  year={2020}
}

@inproceedings{GradNorm,
  title={Gradnorm: Gradient normalization for adaptive loss balancing in deep multitask networks},
  author={Chen, Zhao and Badrinarayanan, Vijay and Lee, Chen-Yu and Rabinovich, Andrew},
  booktitle={International conference on machine learning},
  pages={794--803},
  year={2018},
  organization={PMLR}
}

@inproceedings{DWA,
  title={End-to-end multi-task learning with attention},
  author={Liu, Shikun and Johns, Edward and Davison, Andrew J},
  booktitle={Proceedings of the IEEE/CVF conference on computer vision and pattern recognition},
  pages={1871--1880},
  year={2019}
}

@inproceedings{DTP,
  title={Dynamic task prioritization for multitask learning},
  author={Guo, Michelle and Haque, Albert and Huang, De-An and Yeung, Serena and Fei-Fei, Li},
  booktitle={Proceedings of the European conference on computer vision (ECCV)},
  pages={270--287},
  year={2018}
}

@article{PCGrad,
  title={Gradient surgery for multi-task learning},
  author={Yu, Tianhe and Kumar, Saurabh and Gupta, Abhishek and Levine, Sergey and Hausman, Karol and Finn, Chelsea},
  journal={Advances in neural information processing systems},
  volume={33},
  pages={5824--5836},
  year={2020}
}

@article{GradVec,
  title={Gradient vaccine: Investigating and improving multi-task optimization in massively multilingual models},
  author={Wang, Zirui and Tsvetkov, Yulia and Firat, Orhan and Cao, Yuan},
  journal={arXiv preprint arXiv:2010.05874},
  year={2020}
}

\appendix
\section{Details of LLM-
powered relationship-content alignment}

We employ a multi-modal large language model (LLM) for analysis. The specific prompt used is provided below in Figure \ref{fig:prompt}. The model generates a recommended relationship type for the video.

\begin{figure}[H]
  \centering
\includegraphics[width=0.66\linewidth]{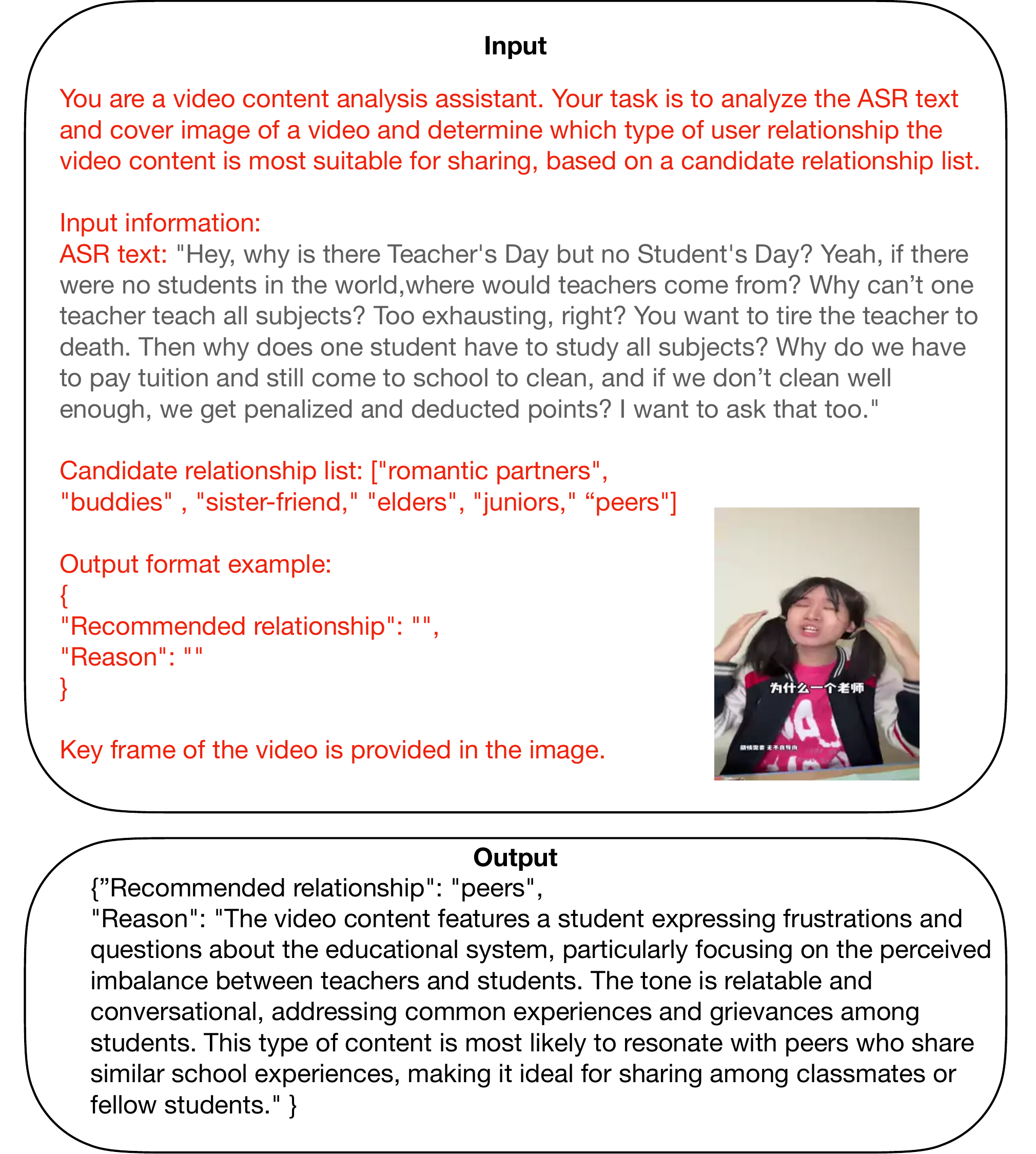}
  \caption{Example Prompt for Relationship-Content Alignment}
  \label{fig:prompt}
\end{figure}

\end{document}